# Inter organizational System Management for integrated service delivery: an Enterprise Architecture Perspective

Abir ELMIR[1], Badr ELMIR[2] and Bouchaib BOUNABAT[3]

Al-Qualsadi Research & Development Team
Ecole Nationale Supérieure d'Informatique et d'Analyses des Systèmes ENSIAS,
Université Mohammed V, Rabat Maroc

**Abstract**

Service sharing is a prominent operating model to support business. Many large inter-organizational networks have implemented some form of value added integrated services in order to reach efficiency and to reduce costs sustainably. Coupling Service orientation with enterprise architecture paradigm is very important at improving organizational performance through business process optimization. Indeed, enterprise architecture management is increasingly discussed because of information system role as part of achieving the strategic direction of value creation and contribution to economic growth. Also, system architecture promotes synergy and business efficiency for inter-organizational collaboration. For this purpose, this work proposes a review of service oriented enterprise architecture. This review, enumerates several integrative and collaborative frameworks for integrated service delivery.

*Keywords:* Integrated delivery network, Information system management, Service orientation, Enterprise architecture framework.

## 1. Introduction

To operate effectively, organizations are encouraged to enter into close interaction with all their partners. Inter organizational collaboration involves an increasing trend for several information systems to span boundaries between organizations. To integrate business processes, organizations have to plan efficiently interaction between systems components of partners.

Enterprise Architecture (EA) can be used to facilitate the integration and to plan effectively the complex inter organizational information system (IOS). Also, Service oriented architecture (SOA) is an architectural paradigm structuring interconnection of distributed systems. SOA aims to successfully integrate existing systems and to create innovating services for customers.

Therefore, in order to manage inter organizational systems for integrated delivery services, this paper proposes a review of integrative and collaborative enterprise architecture frameworks supporting service orientation paradigm.

The paper is structured as follows. The next section introduces the context of integrated delivery networks. It enumerates several requirements to succeed collaboration and deliver value added integrated services. Section 3 is in relation with inter-organizational systems architecture. This section distinguishes different terminology elements used to describe IOS architecture. It reminds the essential IOS implementation means with a specific focus on process driven services. The section 4 proposes a categorization for frameworks to plan IOS for IDN. It notes the importance to explicitly support service orientation and proposes an inventory of integrative and collaborative frameworks to be used for service oriented enterprise architecture (SOEA). This is followed by conclusion in Section 5.

## 2. Integrated Delivery Network

Organizations are more and more information intensive entities. Information technology is central to Integrated Delivery Network (IDN) establishment that enables many organizations to cooperate and allows the sharing of data and services across disparate applications and systems. In IDN context, Interoperability characterizes the ability, for any number of processing information systems, to interact and exchange information and services [1]. Interoperability has become now one of the major concerns of information systems managers. Interoperability gets more challenging in IDN since sharing information and services is so complex. The achievement of interoperability among partners has both technical and organizational aspects. Also, inter organizational interoperability is concerned with defining common goals, modeling integrated business processes and facilitating the collaboration of participants that wish to exchange information. These organizations may have different internal structures and processes.

## 2.1 Collaborative basis for IDN

IDN enables the exchange of information and services between and within organizations. IDN enables electronic collaboration among entities exchanging data by less reliable or less timely means, or among entities who wish to establish the exchange of information [2]. IDN should coordinate inter organizational processes and manage operations throughout a large network of community and shared resources.

The reasons for exchanging data and invoking authorized partners services are many and varied. For the case of healthcare IDN, it includes: (i) informing patient of care decisions, (ii) following up quality of care, (iii) determining if treatments were necessary and reasonable for the purposes of making payments, (iv) responding to healthcare emergencies such as public health threats, (v) performing studies of population health, (vi) conducting research into the effectiveness of existing and emerging treatment mechanisms [3].

In this context, patient-centered healthcare systems involve advanced interactions between: (i) patients (ii) healthcare providers (hospitals, clinics, physicians, public health providers, specialists), (iii) independent laboratories, (iv) community pharmacies, (v) public health agencies (local, regional and national), (vi) pharmaceutical and medical device manufacturers, (vii) researchers (academic, government, and independent), (viii) payers (government or private insurers)[3].

A large number of IDNs are based on partners information systems interoperability rather than the adoption of a unique fully integrated inter organizational system across the network. IDN establishment entails the compliance of involved partners systems to a minimum of information standards before the data exchange interfacing, the service oriented interaction implementation or the composition of new inter organizational business processes.

## 2.2 Quality requirement for collaboration success

Organizations need to develop agility to move in ever-changing contexts. They must overcome a series of challenges in order to establish and sustain cooperation with their partners [4-5, 28-30].

IDN provides a viable environment for collaborative entities allowing them to organize a performance improvement goal. The preparation of IDN is justified, in part, by a series of benefits. However, a set of constraints and challenges accompany the success of business operations. These challenges can be classified in three categories [6-8]. The first one is about functional challenges. The two other categories are related to change requests issues. Indeed, the second class is more interested into context dependent adaptation requests. The third class is more sensitive to requests evolution over time.

**Organizational and functional requirements**

The first class of challenges is Functionality. This class refers to the essential purpose of involved information systems and their components. Functionality capabilities are mainly recognized in requirements identification stage. This class contains various features among which [6-8]:

- Maintaining actors autonomy ;
- Elevating interactions quality with partners ;
- Managing Security risks ;
- Ensuring regulatory compliance ;
- Develop horizontal alignment with IDN Partners.

The second and third classes are related to quality requirements linked to system change management. Change requests can be classified into two main categories:

(i) "Adaptability category" including context dependent change requests and (ii) "evolution category" time dependent change requests.

**Adaptation requirements**

The former category entitled "Adaptability" comprises context dependent change requests includes [6-8]:

- Reusing solutions in new contexts ;
- Simultaneous Existence of resources and complementary services between entities ;
- Ability to renew procedures following IDN membership ;
- Flexibility on change management ;
- Offering service variants depending on use.

**Evolutionary requirements**

The latter category, named "Evolution", includes time dependent change requests and it encloses capabilities like [6-8]:

- Implementing continuous organizational changes;
- Maintaining inter organizational systems ;
- Stabilizing the established environment ;
- Elevating the verification and validation maturity ;
- Scaling solutions following the IDN extension and its development.

# 3. Inter-organizational system for IDN

The concept of enterprise architecture (EA) attracted a lot of interest during the past decade [24]. It aims to provide a structure for business processes and systems that supports them. EA represents information systems using models in order to illustrate interrelationship between their components and the relationship with their ecosystems.

EA proposes to take an inventory of information system components by considering: (1) organization procedures, etc. (2) business process (3) IT applications, (4) technical infrastructure. Indeed, most businesses around the world have established Enterprise Architecture programs [25]. They aim to eliminate overlapping projects, to support reuse, and to enhance interoperability.

On the other hand, several tactical plans were limited to the single issue of cooperation and many interoperability frameworks were developed. They mainly address technical problems by referencing the main recommended specifications to facilitate and promote cooperation within and between organizations [26]. In this sense and in order to facilitate interoperation within a business collaboration network, usually IDN members tend to adopt enterprise architecture as strategic choice of organization using "the service oriented" paradigm and techniques to implement and deploy services.

It exists two approaches in studying inter-organizational systems, as depicted in Figure 1 below [9].

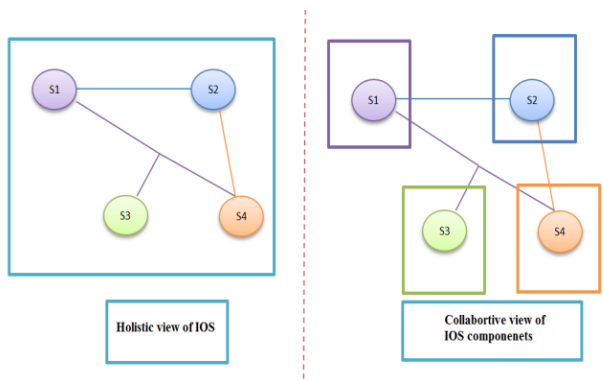

Fig. 1. Holistic versus collaborative views of IOS

The first approach is to perceive the network in a holistic and systemic manner. The network is supported by a single IOS built from the aggregation of IDN members systems [10]. The paradigms of integration and aggregation mechanisms are managing relationships between distributed information systems. The inter-organizational system is managed more evenly [11]. Tools and mechanisms of governance and management are widely shared by different stakeholders. The partners system assembly is better mastered [12].

The second scenario is to focus on the internal partners systems, and then, in a second phase, to foster systems interactions. Most management efforts are internal to each IDN members. Particular importance is attributed to inter-system interactions [13]. The interoperation paradigm is used to manage the partner's relationship [9]. Internal characteristics of each IDN member and the heterogeneity of the solutions are developed and taken into account for IDN adaptation and its evolution [14].

## 3.1 Relations of inter-organizational systems components

Components Cooperation and collaboration of inter-organizational system (IOS) are characterized by information processing capacity to connect and exchange information and services [15]. This ability to interoperate is thus identified as a functional requirement of any computer system to operate and interact with its ecosystem [15, 16].

Interoperability is often confused with other quality concepts [17]. Yet it is quite different for the following terms: compatibility, integration, internal interoperability, data exchange, uniformity and implementation means. In the following, the main difference points between the existing concepts, are illustrated in Figure 2 below and then explained and detailed thereafter.

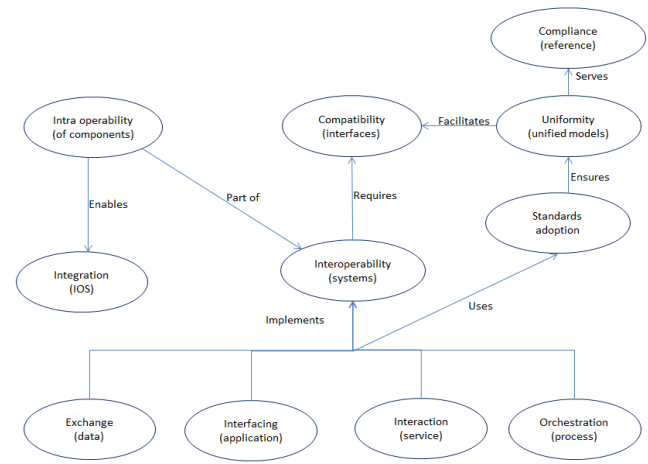

Fig. 2. Architectural collaboration concepts of IOS

**Compatibility**

Interfaces compatibility is a mandatory condition for systems interaction. Interoperability can be seen as a "compatibility sought" to cooperate [18]. IOS operational performance exceeds compatibility requirements of communication interfaces to consider also the implementation details of links as well as quality of interactions.

**Integration**

Integration implies the existence of a unique system, while interoperability requires at least two systems to interact. Systems integration is an ideal approach to reduce barriers impeding collaboration. Also, the maintenance of existing subsystems and the development of interoperability between them remains an implementation means for integrated system solution [19]. However, the independence of governance areas invited to work as a brake for this approach. Interoperability is a characteristic that describes IOS involved subsystems while integration rather qualifies the hole IOS.

**Internal interoperability**

The concept of intra-operability or internal interoperability is the ability of system components to operate with the other components of the same system. Integration is a feature of the overall system but intra operability qualifies internal components in their interactions [20].

**Uniformity**

Standardization aims to ensure that all stakeholders are consistent with a model in order to share common characteristics and facilitate subsequent interoperability. Systems uniformity reduces uncontested incompatibility issues. Interoperability between two systems does not necessarily require compliance with a unified model [21].

**Compliance**

Compliance implies the existence of a prior agreement of stakeholders to adopt the techniques to use. Compliance is ensured through standardization efforts. Compliance significantly reduces organizational and technical barriers. Interoperability does not necessarily require the compliance of systems to a common reference [21-22].

3.2 inter-organizational system implementation means

IOS engineering is done in direct connection with EA levels. This is based on: (i) Infrastructure pooling, (ii) data exchange, (iii) service invocation, (iv) process composition or (v) application integration.

**Infrastructure pooling**

This approach replicates data and services between IDN members' remote sites or shares infrastructure between partners. Pooling makes easy the support of applications, information and pooled services.

**Data exchange interfacing**

This federated approach exchanges data using point-to-point custom interfaces between information subsystems [22-23]. In spite of negative aspects of this approach such as perpetual changes in involved subsystems and interfaces maintenance costs, mainly in large IDN, it is, in many cases, a unique way to exchange data and establish interoperability.

**Service oriented interaction**

In addition to the exchange of information, there is a need for service integration and application reuse. Information systems must be able not only to access and use the services provided by others, but also to reuse their functionality. In this way, it is theoretically possible to build complex information services from the composition of existing ones. This federated approach is used to establish new composite process-oriented services through IDN. It reuses existing services within entities to provide high value-added business services.

**Process composition**

Process composition is concerned with automated means for constructing business processes in IDN [21]. Collaboration among inter organizational processes can be supported by linking the underlying sub-supporting systems that are responsible for executing the corresponding sub-processes within each IDN member. This federated approach is used to establish new composite process-oriented services across IDN. An integrated business process is operationalized in a workflow that can be supported by workflow management technology.

**Standards adoption**

This unified approach suggests the compliance of IOS with a set of commonly accepted data representation and communication standards. As a result, a considerable number of standards have been developed by various organizations [21-23]. Although, standardization is important in this domain, the need for standardization is not sufficient to establish an enduring interoperability as these standards change over time and there are some entities that may not accept these standards in their information processing.

3.3 business process driven service integration

Service-oriented interaction model implements less coupled connections between various distributed software components. The approach seeks to provide abstraction by encapsulating functionality and allowing reuse of existing services. One of the promises of SOA is to compose different functionality exposed as services to produce a high-value business process business perspective. It offers the ability to integrate third-party solutions and easily adapt to new requirements of the trade. Several works deal interference between SOA and EA uses and mention the relationship between these two concepts. Both are independent of technology and requiring similar strategies and planning activities [31]. Authors of [32] highlight the interconnections and differences between these two disciplines in the context of development of electronic government (e-government) arguing that they are to be treated not as alternatives but rather in parallel and combined. Indeed, the adoption of an EA approach greatly facilitates the preparation of solutions based on SOA. Similarly, the SOA extent can cover the different layers of the EA.

In this case, an automated business process, as designed in Figure 3, exposes to its clients a set of business services. This process may be elementary or composite. Composite processes are composed by a set of processes. An elementary process ensures a set of activities. Theses automated activities use IT applications via application services. An integrated business process may be located within a single organization or across organizational boundaries. In this context, clients expect to perceive business as a homogeneous and coherent unit in order to have a unified access to services they need. So, IDN should be prepared to interact effectively with all the surrounding actors. This requires essentially openness and willingness to break functional, organizational and technological barriers. A business process is a set of related activities or operations which, together, create value and assist organizations to achieve their strategic objectives. A systematic focus on improving processes can therefore have a dramatic impact on the effective operation of agencies.

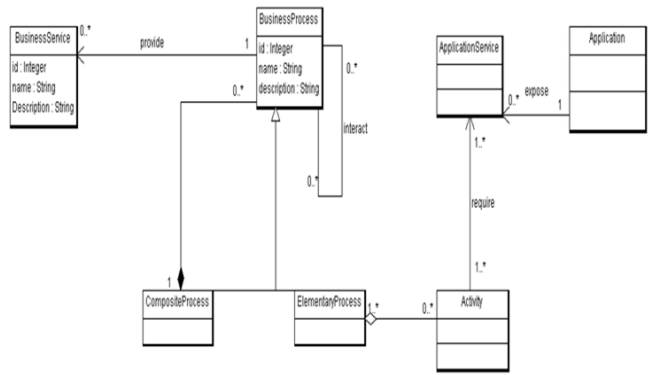

Fig. 3. Model of Business Process Driven Services [37]

Also, there is an increasing trend for several information systems to span boundaries between organizations. Such systems can be used to support collaborations and partnerships among organizations for competitive purposes. Low quality level of inter organizational systems is a potential failure of cooperation and collaboration [27].

Within a collaborative ecosystem, ISQ improvement deals with conceptual, organizational and technical barriers between stakeholders that may belong to different governance subdomains [28].

## 4. Enterprise architecture frameworks for inter-organizational information systems

EAF represent a set of models, implementation methods, working tools and frameworks to facilitate Enterprise Architectures implementation. EAF provide best practices based on successful real experiences in practice. These frameworks are based on different views of the enterprise including: business, applications, technology, infrastructure, etc. The common elements characterizing these architectural frameworks are: business components modeling principles and also methodology to implement them.

Most Enterprise Architecture frameworks (EAF) showcase service orientation. As examples: the framework ARIS [33], used for business process management, provides extended support for Service-oriented modeling as part ArchiMate [34] uses service entities on different levels abstraction and views within the framework. SOEA, defined as enterprise architecture with service orientation style of target architecture [31, 32]. This new "service" layer highlights business services published by business applications. This "business service" is to be distinguished from the "IT service" layer relative, meanwhile, software and services whose technical aspects are exposed by the computer components involved. This description is illustrated in Figure 4 below.

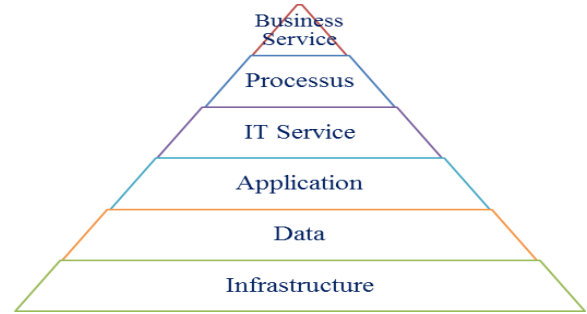

Fig. 4. Service oriented enterprise architecture layers

### 4.1 Integrative EAF for inter-organizational systems

Actually, "service orientation" aims to increase business process automation and to provide more agility for their interconnection. The coupling of service architectural guiding with those of the Enterprise Architecture (EA) is increasingly used for the IS organization in collaboration context. This is federated to a part of collaborative network IDN to give birth to the context of service oriented enterprise architecture (SOEA).

This section describes the concept of Enterprise Architecture in its generic framework and in terms of frameworks available for its implementation. It also pursues the specific SOEA context used increasingly for automation of IDN shared activities.

The "architecture" term can have several meanings depending on the context usage. This term refers to "the art and science of designing buildings and (some) non-building structures" [35]. This generic term exceeds its original domain to reach a wide range of disciplines including systems engineering. Thus the architecture refers to "the abstract representation of the different parts of the system that allows global decisions and ensures the relevance of the assembly, including the consistency and technical efficiency." [13].

The 1471-2000 ANSI/IEEE standard defines the architecture concept as a "the fundamental organization of a system, embodied in its components, their relationships to each other and the environment, and the principles governing its design and evolution" [36].

Figure 5 below summarizes the classification frameworks based on set priorities before detailing this classification thereafter.

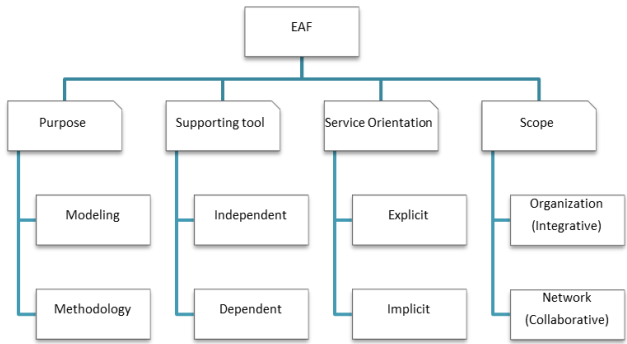

Fig. 5. Classification axis for EAF

The main characteristic of EAF is their high level of abstraction and concentration on modeling aspects. Indeed, most EAF do not explain the methodology to pass from a current architecture state to a target state. This passage remains open and is not detailed in a well-documented process such as ADM approach (Architecture Development Method) of TOGAF framework for example.

Most EAF focus on this transition aspect between the current state and the target state, the majority of EAF are restricted to representation aspects by the prospects they want to value without explicit and equip the architectural evolution. According to the differentiation axes mentioned before, the integrative EAF can be classified as depicted on Table 1.

Table 1: integrative EAF for IDN [37]

| EAF | Purpose | Tool | Service orient. |
|---|---|---|---|
| **Zachman** [38] | Modeling | Independent | Implicit |
| **ARIS** [39] | Modeling | ARIS IT Architect | Explicit |
| **GERAM** [40] | Methodology | Independent | Implicit |
| **SOM** [41] | Modeling | SOM environment | Implicit |
| **EAMIT** [42] | Methodology | Independent | Implicit |
| **EATUL** [43] | Modeling | Independent | Explicit |
| **Archimate** [44] | Modeling | Independent | Explicit |
| **EAKTH** [45] | Modeling | EA Tool | Implicit |
| **FEAR** [46] | Methodology | Independent | Implicit |
| **DEMO** [47] | Methodology | Independent | Implicit |
| **EA³** [37] | Methodology | Independent | Implicit |
| **DYA** [48] | Methodology | Independent | Explicit |
| **Niemann** [49] | Methodology | Independent | Explicit |
| **SAP EAF** [37] | Modeling | SAP EAF toolset, ARIS IT Architect | Explicit |
| **IAF** [50] | Modeling | Independent | Explicit |

Most EAF are supported by modeling tools that allow the enterprise architecture establishment and its maintenance. As examples, there is a plugin for the Eclipse integrated development environment to support different views of TOGAF. ARIS is supported by the "ARIS Toolset" environment. It remains to note that the majority of commercial EAF are restricted to the use of tools for publishers of these EAF or integration projects led by some design offices. The last line of differentiation between EAF is the explicit consideration of service paradigm. Indeed, service orientation principle is stated explicitly in many EAF. The remaining EAF, although they do not explicitly specify the service layer in their meta-models, may support the concept of service as a special layer when mapping the information system. Several works are proposed precisely to shed light on how to integrate the concept of service in the most popular frameworks such as TOGAF or Zachman for example, and while proposing several alternatives to this end [51].

Another line of differentiation is in charge taking perspective of collaborative EAF. Indeed, most of the EAF are designed to be used within a single organization. These EAF can characterize the IOS as a whole by taking it as a single IF. These are called integrative EAF. A second category of collaborative EAF sits there. These frameworks are designed specifically to describe the relationships between IDN member's subsystems. The next section illustrates the characteristics of this category and describes some of these frameworks.

## 4.2 Collaborative EAF for inter-organizational systems

Service oriented enterprise architecture promotes the establishment and automation of IDN inter-organizational processes. SOEA adoption reduces the investment needed to work with the partners. This is possible especially with the availability of lending services to be composed and orchestrated in order to compose new inter-organizational macro processes [4].

Furthermore, having services with a standard and interpretable description allows their discovery and invocation automatically and dynamically what prepares the conditions for establishment of inter-organizational cooperation scenarios on demand [5].

The evolution of service orientation is marked by the exhibition, in addition to business and technical services, new types of services focused on architectural elements not covered before. This is data services and infrastructure services essentially blown by two trends that are open data and cloud services (pooling of new types of services) [52]. The guidance service coupled with the classical model of Enterprise Architecture in figure 6 below:

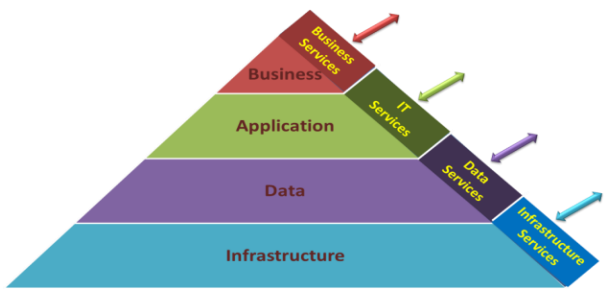

Fig. 6. Service orientation for different architectural layers

The EA used to describe both the internal interconnections organizations and those external to IDN. It also helps to plan possible changes at the organizational level and at the collaboration support systems level. To differentiate internal and external coverage levels, the concept of Extended Enterprise Architecture (Extended Enterprise Architecture EEA) is introduced. This same model extended to IDN scale can be represented as follows (see figure 7 below) :

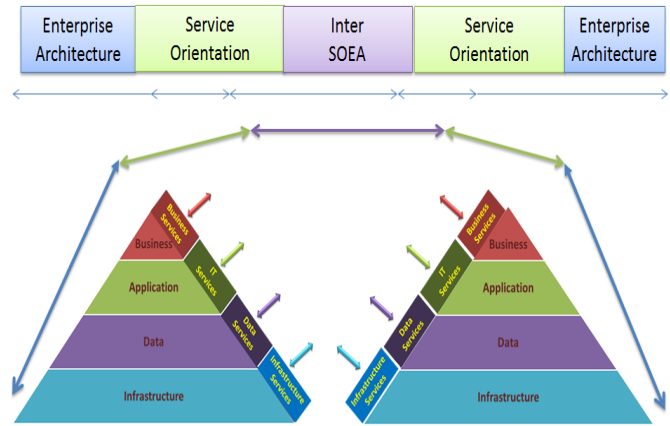

Fig. 7. Extended service oriented enterprise architecture over IDN

Few EAF explicitly take account of collaborative environments in which organizations are required to participate. Among these frameworks: CFCEBPM (Collaboration Framework for Cross-enterprise Business Process Management) [61], ARDIN EVEI (ARDIN Extension for Virtual Enterprise Integration) [58], VECCF (Virtual Enterprise Chain Collaboration Framework)[64] and E2AF (Extended Enterprise architecture framework) [59] or ARCON (Architecture for Collaborative Network of Reference) [66].

If the proposals of [61] and [58] and [66] provide a methodology for the implementation of the collaboration architecture, the other frameworks are essentially concerned with the evolution between the intermediate states of information systems (ie As-Is to to-be). The method of [61] consists of five phases dealing firstly common collaborative process as well as local processes in each. Regarding the methodology proposed by [58], its structure is quite similar to that of [61] although it gives more importance to the evolution As-Is to the to-Be.

Table 2: collaborative EAF for IDN

| EAF | Intérêt | Outil | Orient. service |
|---|---|---|---|
| **MEMO** [53] | Modeling | Independent | Implicit |
| **DoDAF** [54] | Modeling | Independent | Implicit |
| **FEAF** [55] | Modeling | Independent | Explicit |
| **SEAM** [56] | Modeling | Independent | Explicit |
| **Gallen** [57] | Methodology | ADOben | Explicit |
| **ARDIN-EVEI** [58] | Methodology | Independent | Explicit |
| **E2AF** [59] | Methodology | Independent | Explicit |
| **BEAMS** [60] | Modeling | Independent | Explicit |
| **CFCEBPM** [61] | Methodology | ARIS IT Architect | Explicit |
| **MoDAF** [62] | Modeling | Independent | Implicit |
| **SAGA** [63] | Modeling | Independent | Explicit |
| **VECCF** [64] | Modeling | Independent | Explicit |
| **TOGAF** [65] | Methodology | Plugin for Eclipse | Explicit |
| **ARCON** [66] | Modeling | Independent | Explicit |
| **Hanschke** [67] | Methodology | iteraplan | Explicit |

ARDIN-EVEI, CFCEBPM and ARCON provide different modeling language from each other [58, 61, 66]. For CFCEBPM [61], it is necessary to use tools for visualizing the collaborative process and ensure a common understanding of the collaborative process between all the entities involved in the collaborative process. They propose to use a specific software (based on ARIS IT Architect) using the modeling language BPML (Business Process Modeling Language). The authors of VECCF [64] propose to use a neutral design platform architecture based models (Model Driven Architecture - MDA). This platform is based on UML (Unified Modeling Language). Work [58], meanwhile, proposes the use of IDEF and GRAI to represent a general level of various activities and decisions different companies. They also propose to use UML to describe the process automated trades.

## 5. Conclusions

Service oriented enterprise architecture (SOEA) is used to plan and control the construction of systems. This discipline provides models to understand how the parts of the enterprise fit together. It processes a variety of issues such as business agility, flexibility, interoperability, alignment or governance. SOEA is used increasingly to manage collaboration especially if the service orientation is supported as a structuring choice. The adoption of service orientation paradigm to govern collaborative situations represents a framework for inter-organizational processes integration. This is a prerequisite adopted by several IDN precisely to promote and better manage collaboration between partners and to share architectural visions. Network operation and organization should be structured and modeled via architectures designed to support inter-organizational processes and the integration of partner information systems. In this sense, some frameworks are proposed to frame collaboration and meet implementation requirements (framework, methodology/modeling language, supported tools). The present paper enumerates two categories of these frameworks: integrative EAF to better master IOS management and collaborative EAF to have more agility for handling adaptation and evolution requests across IDN.

**A. ELMIR** "Ph.D. candidate" at ENSIAS (National Higher School for Computer Science and System analysis) Rabat, holder of a master degree from ENIM Rabat and a mastere diploma from INSA Lyon France (2011). Her research focuses on multi objective optimization and optimal control of information system quality within collaborative networks. She is an integration architect on a private Financial Holding (Banking, Insurance). She also runs the Solutions support activity of this holding.

**B. ELMIR** He received a Ph.D. degree (2012), an Extended Higher Studies Diploma (2006) and a Software engineer degree (2002) from ENSIAS, (National Higher School for Computer Science and System analysis), Rabat. His research currently focuses on interoperability optimization on public administration. He is an integration architect on the Ministry of Economy and Finance of Morocco since 2002. He also oversees information system quality assurance activity in this department.

**B. BOUNABAT** Ph.D. in Computer Sciences. Professor in ENSIAS, (National Higher School for Computer Science and System analysis), Rabat, Morocco. International Expert in ICT Strategies and E-Government to several international organizations, Member of the board of Internet Society - Moroccan Chapter.